\documentclass[amsmath,aps,showpacs,a4paper,10pt]{revtex4}
 \usepackage{epsf}
 \usepackage{graphicx}    % Include figure files

\begin{document}

 \newcommand{\be}[1]{\begin{equation}\label{#1}}
 \newcommand{\ee}{\end{equation}}
 \newcommand{\bea}{\begin{eqnarray}}
 \newcommand{\eea}{\end{eqnarray}}
 \def\disp{\displaystyle}

 \def\gsim{ \lower .75ex \hbox{$\sim$} \llap{\raise .27ex \hbox{$>$}} }
 \def\lsim{ \lower .75ex \hbox{$\sim$} \llap{\raise .27ex \hbox{$<$}} }

 \begin{titlepage}

 \begin{flushright}
 arXiv:1108.0859
 \end{flushright}

 \title{\Large \bf Constraining $f(T)$ Theories with the
 Varying Gravitational~Constant}

 \author{Hao~Wei\,}
 \email[\,email address:\ ]{haowei@bit.edu.cn}
 \affiliation{School of Physics, Beijing Institute
 of Technology, Beijing 100081, China}

 \author{Hao-Yu~Qi\,}
 \affiliation{School of Physics, Beijing Institute
 of Technology, Beijing 100081, China}

 \author{Xiao-Peng~Ma}
 \affiliation{School of Physics, Beijing Institute
 of Technology, Beijing 100081, China}

 \begin{abstract}\vspace{1cm}
 \centerline{\bf ABSTRACT}\vspace{2mm}
 As is well known, a varying effective gravitational ``constant'' is
 one of the common features of most modified gravity theories.
 Of course, as a modified gravity theory, $f(T)$ theory is not
 an exception. Noting that the observational constraint on the
 varying gravitational ``constant'' is very tight, in the
 present work we try to constrain $f(T)$ theories with the
 varying gravitational ``constant''. We find that the allowed
 model parameter $n$ or $\beta$ has been significantly shrunk
 to a very narrow range around zero. In fact, the results
 improve the previous constraints by an order of magnitude.
 \end{abstract}


 \maketitle

 \end{titlepage}

 \renewcommand{\baselinestretch}{1.1}

%============================= section 1 ===================================

\section{Introduction}\label{sec1}

Motivated by the well-known large number hypothesis proposed
 in 1937~\cite{r1}, the varying fundamental ``constants'' have
 remained as one of the unfading subjects for decades. In fact,
 the gravitational constant $G$ was the first one whose
 constancy was questioned by Dirac~\cite{r1}.

As a pure gravitational phenomenon, the variation of the
 gravitational ``constant'' does not affect the local physics
 (e.g. the atomic transitions or the nuclear physics), and
 hence most constraints on it are obtained from systems in
 which gravity is non-negligible, such as the motion of the
 bodies of solar system and astrophysical systems~\cite{r2}.
 Following~\cite{r2}, here we briefly review the current
 observational constraints on the varying gravitational
 ``constant''. The first type of constraints arises from
 solar system. The latest analysis of the lunar laser ranging
 experiment~\cite{r3} gives the constraint
 \be{eq1}
 |\dot{G}/G|\leq 1.3\times 10^{-12}\,{\rm yr}^{-1}\,,
 \ee
 where a dot denotes the derivative with respect to the
 time $t$. In~\cite{r4}, the combination of Mariner~10, Mercury
 and Venus ranging data gives
 \be{eq2}
 |\dot{G}/G|\leq 2\times 10^{-12}\,{\rm yr}^{-1}\,.
 \ee
 In~\cite{r5}, the ranging data from Viking landers on Mars
 lead to the constraint
 \be{eq3}
 |\dot{G}/G|\leq 6\times 10^{-12}\,{\rm yr}^{-1}\,.
 \ee
 The second type of constraints arises from pulsar timing.
 In~\cite{r6}, the constraints from PSR~B1913+16
 and PSR~B1855+09 are given by
 \be{eq4}
 |\dot{G}/G|\leq 9\times 10^{-12}\,{\rm yr}^{-1}\,,
 \ee
 and
 \be{eq5}
 |\dot{G}/G|\leq 2.7\times 10^{-11}\,{\rm yr}^{-1}\,,
 \ee
 respectively. On the other hand, the constraint from
 PSR~J0437-4715 reads~\cite{r7}
 \be{eq6}
 |\dot{G}/G|< 2.3\times 10^{-11}\,{\rm yr}^{-1}\,.
 \ee
 The third type is the stellar constraints. In~\cite{r8}, the
 ages of globular clusters give the constraint
 \be{eq7}
 |\dot{G}/G|\leq 3.5\times 10^{-11}\,{\rm yr}^{-1}\,.
 \ee
 The constraint from helioseismology is given by~\cite{r9}
 \be{eq8}
 |\dot{G}/G|< 1.6\times 10^{-12}\,{\rm yr}^{-1}\,.
 \ee
 The seismology of white dwarf G117-B15A~\cite{r10} gives
 the constraint
 \be{eq9}
 |\dot{G}/G|< 4.1\times 10^{-11}\,{\rm yr}^{-1}\,.
 \ee
 In~\cite{r11}, the constraint from the cooling of white
 dwarfs reads
 \be{eq10}
 |\dot{G}/G|< 2\times 10^{-11}\,{\rm yr}^{-1}\,.
 \ee
 In~\cite{r12}, the light curves of supernovae give the constraint
 \be{eq11}
 |\dot{G}/G|\leq 4.8\times 10^{-12}\,{\rm yr}^{-1}\,.
 \ee
 Finally, one can also obtain the cosmological constraints on
 the varying $G$ from cosmic microwave background (CMB) and big
 bang nucleosynthesis (BBN). However, this type of cosmological
 constraints on the varying $G$ heavily relies on not only the
 assumption of the whole history of $G(t)$ but also many other
 complicated factors. Therefore, it is very difficult to
 state a definitive constraint~\cite{r2}. So, here we do not
 consider this type of the cosmological constraints on the
 varying gravitational ``constant''.

Obviously, the currently tightest observational constraint on
 the varying gravitational ``constant'' is the one given in
 Eq.~(\ref{eq1}),
 namely, $|\dot{G}/G|\leq 1.3\times 10^{-12}\,{\rm yr}^{-1}$.
 So, in the present work we only use this tightest
 observational constraint given in Eq.~(\ref{eq1})
 to constrain $f(T)$ theories.

In fact, it is well known that a varying effective gravitational
 ``constant'' is one of the common features of many modified
 gravity theories~\cite{r2}, for examples, $f(R)$ theory,
 scalar-tensor theory (including Brans-Dicke theory and
 Galileon theory), braneworld scenarios (such as DGP, RSI and
 RSII), $f(\cal G)$ theory ($\cal G$ is the Gauss-Bonett term),
 Horava-Lifshitz theory, MOND and TeVeS theories (see
 e.g.~\cite{r13,r14,r15,r16,r17} for reviews). Recently, a new
 modified gravity theory, namely the so-called $f(T)$ theory,
 attracted much attention in the community, where $T$ is the
 torsion scalar. Similar to many modified gravity theories,
 the effective gravitational ``constant'' is also varying in
 $f(T)$ theory. Therefore, it is of interest to constrain
 $f(T)$ theory with the varying gravitational ``constant''.
 This is what we need to do in the present work.

In Sec.~\ref{sec2}, we briefly review the key points of
 $f(T)$ theory. In Sec.~\ref{sec3a}, in order to compare
 with the observational constraint on the varying gravitational
 ``constant'', we give the corresponding formula of
 $|\dot{G}_{\rm eff}/G_{\rm eff}|$ for a general $f(T)$
 theory. In Secs.~\ref{sec3b} and~\ref{sec3c}, we constrain
 two concrete $f(T)$ theories with the varying gravitational
 ``constant'', respectively. Finally, a brief conclusion
 is given in Sec.~\ref{sec4}.

%============================= section 2 ===================================

\section{A brief review of $f(T)$ theory}\label{sec2}

$f(T)$ theory is a generalization of the teleparallel
 gravity originally proposed by Einstein~\cite{r18,r19}. In
 teleparallel gravity, the Weitzenb\"ock connection is used,
 rather than the Levi-Civita connection which is used in
 general relativity. Following~\cite{r20,r21}, here we briefly
 review the key points of $f(T)$ theory. The orthonormal tetrad
 components $e_i(x^\mu)$ relate to the metric through
 \be{eq12}
 g_{\mu\nu}=\eta_{ij}e_\mu^i e_\nu^j\,,
 \ee
 where Latin $i$, $j$ are indices running over 0, 1, 2, 3 for
 the tangent space of the manifold, and Greek $\mu$,~$\nu$ are
 the coordinate indices on the manifold, also running over 0,
 1, 2, 3. In $f(T)$ theory, the gravitational action is given by
 \be{eq13}
 {\cal S}_T=\frac{1}{2\kappa^2}\int d^4 x\,|e|\,
 \left[\,T+f(T)\,\right]\,,
 \ee
 where $\kappa^2\equiv 8\pi G_N$ ($G_N$ is the Newton constant), and
 $|e|={\rm det}\,(e_\mu^i)=\sqrt{-g}\,$. It is worth noting that in
 the literature, $T+f(T)$ in Eq.~(\ref{eq13}) could be instead
 replaced by $f(T)$, and hence one should be aware of the
 correspondence between these two formalisms. The torsion
 scalar $T$ is defined by
 \be{eq14}
 T\equiv{S_\rho}^{\mu\nu}\,{T^\rho}_{\mu\nu}\,,
 \ee
 where
 \bea
 {T^\rho}_{\mu\nu} &\equiv &-e^\rho_i\left(\partial_\mu e^i_\nu
 -\partial_\nu e^i_\mu\right)\,,\label{eq15}\\
 {K^{\mu\nu}}_\rho &\equiv &-\frac{1}{2}\left({T^{\mu\nu}}_\rho
 -{T^{\nu\mu}}_\rho-{T_\rho}^{\mu\nu}\right)\,,\label{eq16}\\
 {S_\rho}^{\mu\nu} &\equiv &\frac{1}{2}\left({K^{\mu\nu}}_\rho
 +\delta^\mu_\rho {T^{\theta\nu}}_\theta-
 \delta^\nu_\rho {T^{\theta\mu}}_\theta\right)\,.\label{eq17}
 \eea
 In this work, we consider a spatially flat
 Friedmann-Robertson-Walker (FRW) universe. In this case, one
 easily finds that~\cite{r20,r21}
 \be{eq18}
 T=-6H^2,
 \ee
 where $H\equiv\dot{a}/a$ is the Hubble parameter; $a=(1+z)^{-1}$ is
 the scale factor (we have set $a_0$=1, whereas the subscript
 ``$0$'' indicates the present value of the corresponding quantity);
 $z$ is the redshift. Therefore, one can use $T$ and $H$
 interchangeably. The modified Friedmann equation and Raychaudhuri
 equation are given by~\cite{r21,r22,r23}
 \be{eq19}
 H^2=\frac{\kappa^2}{3}\rho-\frac{f}{6}-2H^2 f_T\,,
 \ee
 \be{eq20}
 \left(H^2\right)^\prime
 =\frac{2\kappa^2 p+6H^2+f+12H^2 f_T}{24H^2 f_{TT}-2-2f_T}\,,
 \ee
 where $f_T\equiv\partial f/\partial T$, a prime denotes the
 derivative with respect to $\ln a$, and $\rho$, $p$ are
 the total energy density and pressure, respectively. In an
 universe with only dust matter, $p=p_m=0$ and $\rho=\rho_m$.
 Obviously, if $f(T)=const.$, $f(T)$ theory reduces to the
 well-known $\Lambda$CDM model.

In fact, $f(T)$ theory was firstly used to drive inflation
 by Ferraro and Fiorini~\cite{r24,r25}. Later, Bengochea
 and Ferraro~\cite{r20}, as well as Linder~\cite{r21},
 proposed to use $f(T)$ theory to drive the current accelerated
 expansion of our universe without invoking dark energy. Very
 soon, $f(T)$ theory attracted much attention in the community.
 We refer to e.g.~\cite{r26,r27,r28,r29,r30,r31,r36} for
 relevant work.

%============================= section 3 ===================================

\section{Constraining $f(T)$ theories with the varying
 gravitational~``constant''}\label{sec3}

%============================= section 3.1 ===================================

\subsection{$|\dot{G}_{\rm eff}/G_{\rm eff}|$ for a general
 $f(T)$ theory}\label{sec3a}

As is well known, a varying effective gravitational ``constant'' is
 one of the common features of most modified
 gravity theories~\cite{r2}. Of course, as a modified gravity
 theory, $f(T)$ theory is not an exception. In fact, the effective
 gravitational ``constant'' of $f(T)$ theory has been derived
 in~\cite{r30}, namely
 \be{eq21}
 G_{\rm eff}=\frac{G_N}{1+f_T}\,.
 \ee
 Obviously, if $f(T)$ is a linear function of $T$, namely
 $f(T)=\alpha T$, the effective gravitational constant is just
 rescaled to be $G_{\rm eff}=G_N/(1+\alpha)$, which is still
 constant in time. However, in general $G_{\rm eff}$ is varying
 for any non-linear $f(T)$. From Eq.~(\ref{eq21}), it is easy
 to get
 \be{eq22}
 \left|\,\frac{\dot{G}_{\rm eff}}{G_{\rm eff}}\,\right|=
 \left|\,\frac{f_{TT}\dot{T}}{1+f_T}\,\right|\,.
 \ee
 Using Eq.~(\ref{eq18}) and the universal relation
 $\dot{x}=-(1+z)H(dx/dz)$ for any function $x$, we have
 \be{eq23}
 \dot{T}=-(1+z)\,T_0\,H\,\frac{dE^2}{dz}\,,
 \ee
 where $E^2\equiv T/T_0=H^2/H_0^2$. Substituting $f_T$,
 $f_{TT}$ and Eq.~(\ref{eq23}) into Eq.~(\ref{eq22}) and
 setting $z=0$, the present value of
 $|\dot{G}_{\rm eff}/G_{\rm eff}|$ is on hand (note that the
 observational constrain given in Eq.~(\ref{eq1}) is obtained
 from the lunar laser ranging experiment, at redshift $z=0$).
 It is easy to see that the Hubble constan $H_0$ will appear
 in the final result of $|\dot{G}_{\rm eff}/G_{\rm eff}|$
 through $\dot{T}$ at redshift $z=0$. According to~\cite{r32},
 \be{eq24}
 H_0=100\,h~{\rm km/s/Mpc}=1.02275\times 10^{-10}
 \,h~{\rm yr}^{-1}\,.
 \ee
 Note that we use the units $\hbar=c=1$ throughout this work.
 It is easy to see that the Hubble constant $H_0$ is a suitable
 measurement of the present value of
 $|\dot{G}_{\rm eff}/G_{\rm eff}|$. Very recently, the SHOES
 (Supernovae and H0 for the Equation of State) Hubble Space
 Telescope program~\cite{r33,r34} released
 its latest model-independent measurement of Hubble constant,
 namely $h=0.738\pm 0.024$ (with only $3.3\%$
 uncertainty)~\cite{r34}. Therefore, we adopt $h=0.738$ in
 the following.

Before going further, we would like to say some words here.
 It is worth noting that the $G_{\rm eff}$ given
 in Eq.~(\ref{eq21}) was obtained by using linear
 theory~\cite{r30}. However, the Earth-Moon system is far
 from linearity, at least concerning the density contrast
 (we thank the anonymous referee for pointing out this
 issue). It is well known that via a conformal transformation,
 $f(R)$ theory can be equivalent to General Relativity~(GR)
 with a scalar field (which is coupled with matter) in Einstein
 frame~\cite{r13,r14,r15,r16,r17}. In this case, one finds
 that $f(R)$ theory can evade the local gravity tests through
 the so-called chameleon mechanism~\cite{r35}, in which the
 gravity is effectively screened in the solar system. However,
 this way is not viable in $f(T)$ theory unfortunately. In
 e.g.~\cite{r36}, it is found that $f(T)$ theory cannot be
 equivalent to teleparallel action plus a scalar field via a
 conformal transformation, because an additional
 scalar-torsion coupling term cannot be removed by the
 conformal transformation. This point makes $f(T)$ and $f(R)$
 theories different. As a result, also in~\cite{r36}, it is
 claimed that the chameleon mechanism might not work in
 $f(T)$ theory, and then it might be hard to evade the solar
 system tests, unlike $f(R)$ theory. In fact, we will find
 that $f(T)$ theory is tightly constrained by the solar
 system tests in the following.

In the next subsections, we will consider two concrete $f(T)$
 theories, namely, $f(T)=\mu(-T)^n$
 and $f(T)=-\mu T\left(1-e^{\beta T_0/T}\right)$, which are
 the most popular $f(T)$ theories discussed extensively in
 the literature (see
 e.g.~\cite{r20,r21,r22,r23,r26,r27,r28,r29,r30,r31}).

%==================== table 1 ====================

 \begin{table}[tbp]
 \begin{center}
 \begin{tabular}{cccccc} \hline\hline\\[-4mm]
 Description  & Best fit & $2\sigma$ right edge
  & $2\sigma$ left edge & $2\sigma$ top edge
  & $2\sigma$ bottom edge \\[0.5mm] \hline\\[-3.5mm]
 ~~($\Omega_{m0}$, $n$)~~ & ~~(0.272, 0.04)~~
  & ~~(0.308, 0.04)~~ & ~~(0.240, 0.04)~~ & ~~(0.272, 0.26)~~
  & ~~(0.272, $-0.29$)~~\\[0.7mm] \hline\\[-3.5mm]
 $|\dot{G}_{\rm eff}/G_{\rm eff}|$ & 1.86849 & 2.01124
  & 1.72109 & 15.8991 & 10.1177 \\[0.7mm] \hline\hline
 \end{tabular}
 \end{center}
 \caption{\label{tab1} The present value of
 $|\dot{G}_{\rm eff}/G_{\rm eff}|$ (in units of
 $10^{-12}\,{\rm yr}^{-1}$) with the best-fit parameters
 of~\cite{r22} and the corresponding $2\sigma$ edge for the
 case of $f(T)=\mu(-T)^n$.}
 \end{table}

%=================================================

\vspace{-3mm}  % used here just for a comfortable typesetting

%============================= section 3.2 ===================================

\subsection{$f(T)=\mu(-T)^n$}\label{sec3b}

At first, we consider the case of $f(T)=\mu(-T)^n$, where
 $\mu$ and $n$ are both constants. This is the simplest model,
 and has been considered in most papers on $f(T)$ theory.
 Obviously, if $n=0$, it reduces to $\Lambda$CDM model.
 Substituting $f(T)=\mu(-T)^n$ into the modified Friedmann
 equation~(\ref{eq19}), one easily finds that $\mu$ is not
 an independent model parameter, namely~\cite{r21,r22,r31}
 \be{eq25}
 \mu=\frac{1-\Omega_{m0}}{2n-1}\left(6H_0^2\right)^{1-n}=
 \frac{1-\Omega_{m0}}{2n-1}\left(-T_0\right)^{1-n},
 \ee
 where $\Omega_{m0}\equiv\kappa^2 \rho_{m0}/(3H_0^2)$ is the
 present fractional energy density of dust matter. So, we have
 \be{eq26}
 f(T)=\frac{1-\Omega_{m0}}{2n-1}(-T_0)\left(
 \frac{T}{T_0}\right)^n\,,
 \ee
 and then
 \be{eq27}
 f_T=\frac{n(1-\Omega_{m0})}{1-2n}E^{2(n-1)}\,,~~~~~~~
 f_{T0}=\frac{n(1-\Omega_{m0})}{1-2n}\,,
 \ee
 \be{eq28}
 f_{TT}=\frac{n(n-1)(1-\Omega_{m0})}{1-2n}\frac{E^{2(n-2)}}{T_0}\,,
 ~~~~~~~f_{TT0}=\frac{n(n-1)(1-\Omega_{m0})}{(1-2n)T_0}\,,
 \ee
 where $E^2\equiv T/T_0=H^2/H_0^2$. Substituting
 $f(T)=\mu(-T)^n$ and Eq.~(\ref{eq25}) into the modified
 Friedmann equation~(\ref{eq19}), we find that~\cite{r22,r31}
 \be{eq29}
 E^2=\Omega_{m0}(1+z)^3+(1-\Omega_{m0})E^{2n}\,.
 \ee
 Obviously, if $n=0$, it reduces to the one of $\Lambda$CDM
 model. Differentiating Eq.~(\ref{eq29}) with respect to
 redshift $z$, we obtain
 \be{eq30}
 \frac{dE^2}{dz}=\frac{3\Omega_{m0}(1+z)^2}
 {1-n(1-\Omega_{m0})E^{2(n-1)}}\,.
 \ee

% we insert Fig. 1 here by breaking Eqs. (30) and (31)
% just for a comfortable typesetting
% when one moves Fig. 1 to other place, please remind that
% originally Eqs. (30) and (31) are unbroken

%============================= Fig. 1 =================================

 \begin{center}
 \begin{figure}[tbp]
 \centering
 \includegraphics[width=0.5\textwidth]{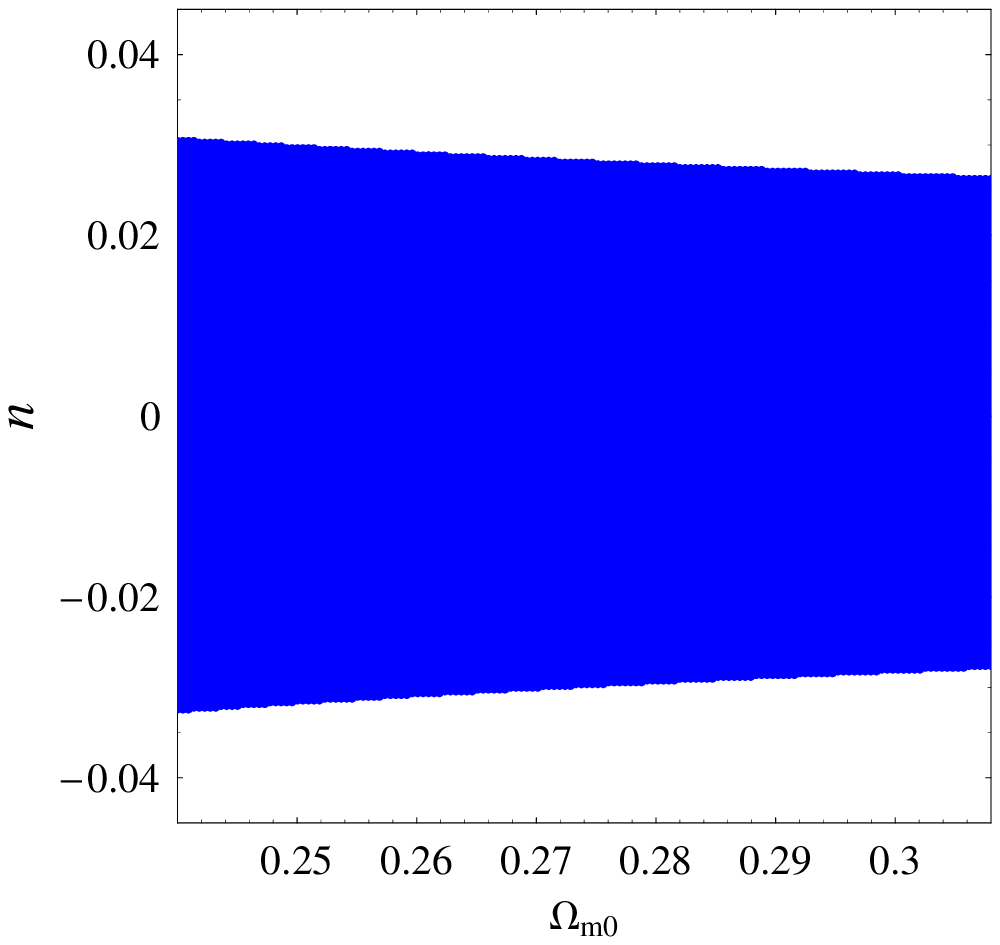}
 \caption{\label{fig1} The viable region in the
 $\Omega_{m0}-n$ parameter space which can satisfy the
 observational constraint on the varying gravitational
 ``constant'' given in Eq.~(\ref{eq1}), as well as the
 latest cosmological data (SNIa+BAO+CMB), for the case
 of $f(T)=\mu(-T)^n$. See text for details.}
 \end{figure}
 \end{center}

%======================================================================

\vspace{-7mm}  % used here just for a comfortable typesetting

\noindent Substituting Eqs.~(\ref{eq27}), (\ref{eq28})
 and~(\ref{eq30}) into Eq.~(\ref{eq23}) and then
 Eq.~(\ref{eq22}), it is easy to see that at redshift $z=0$,
 we have
 \be{eq31}
 \left|\,\frac{\dot{G}_{\rm eff}}{G_{\rm eff}}\,\right|=
 \left|\,\frac{3n(n-1)(1-\Omega_{m0})\Omega_{m0}H_0}
 {(1-n)^2-n^2\Omega_{m0}^2}\,\right|\,.
 \ee
 Therefore, if the model parameters $\Omega_{m0}$ and $n$ are
 given, we can correspondingly get the present value of
 $|\dot{G}_{\rm eff}/G_{\rm eff}|$ from Eq.~(\ref{eq31}).
 Note that in~\cite{r22}, this $f(T)=\mu(-T)^n$ theory has been
 constrained by using the latest cosmological data, i.e., 557
 Union2 type Ia supernovae (SNIa) dataset, baryon acoustic
 oscillation (BAO), and cosmic microwave background (CMB) data
 from WMAP7. The corresponding $2\sigma$ results are given
 by~\cite{r22}
 \be{eq32}
 \Omega_{m0}=0.272^{+0.036}_{-0.032}\,,~~~~~~~
 n=0.04^{+0.22}_{-0.33}\,.
 \ee
 At first, we try to see whether the present value
 of $|\dot{G}_{\rm eff}/G_{\rm eff}|$ with the best-fit
 parameters of~\cite{r22} and the corresponding $2\sigma$ edge
 can satisfy the observational constraint on the varying
 gravitational ``constant'' given in Eq.~(\ref{eq1}), namely,
 $|\dot{G}/G|\leq 1.3\times 10^{-12}\,{\rm yr}^{-1}$. From
 Table~\ref{tab1}, it is easy to see that {\em none} of them
 can satisfy the observational constraint on the varying
 gravitational ``constant'' given in Eq.~(\ref{eq1}). Also from
 Table~\ref{tab1}, we find that for a fixed $n$, the smaller
 $\Omega_{m0}$, the smaller $|\dot{G}_{\rm eff}/G_{\rm eff}|$
 is, whereas $|\dot{G}_{\rm eff}/G_{\rm eff}|$ is also somewhat
 correlated with the extent of $n$ deviating from zero.
 Next, we try to find the viable region in the $\Omega_{m0}-n$
 parameter space which can satisfy the observational constraint
 on the varying gravitational ``constant'' given
 in Eq.~(\ref{eq1}), as well as the latest cosmological data
 (SNIa+BAO+CMB). To this end, we scan the $\Omega_{m0}-n$
 parameter space within the $2\sigma$ region of~\cite{r22},
 namely $0.240\leq\Omega_{m0}\leq 0.308$ and $-0.29\leq n\leq 0.26$
 (see Eq.(\ref{eq32})), and correspondingly calculate the
 present value of $|\dot{G}_{\rm eff}/G_{\rm eff}|$ at every
 scanned ($\Omega_{m0}$, $n$) point. The viable parameter
 region is determined by $|\dot{G}_{\rm eff}/G_{\rm eff}|\leq
 1.3\times 10^{-12}\,{\rm yr}^{-1}$, and we present it in
 Fig.~\ref{fig1}, which is approximately a trapezoid region. It
 is easy to see that the allowed $\Omega_{m0}$ is still
 unchanged, i.e., $0.240\leq\Omega_{m0}\leq 0.308$. However,
 the allowed $n$ is significantly shrunk from the original
 $-0.29\leq n\leq 0.26$ of~\cite{r22} to a very narrow range
 around zero, namely
 \be{eq33}
 -0.032\;\lsim\; n\;\lsim\; 0.030\,.
 \ee
 This is the additional constraint on $f(T)=\mu(-T)^n$ theory
 from the varying gravitational ``constant''. Clearly, this
 result improves the previous constraint by an order of
 magnitude.

%==================== table 2 ====================

 \begin{table}[bp]
 \begin{center}
 \begin{tabular}{cccccc} \hline\hline\\[-4mm]
 Description  & Best fit & $2\sigma$ right edge
  & $2\sigma$ left edge & $2\sigma$ top edge
  & $2\sigma$ bottom edge \\[0.5mm] \hline\\[-3.5mm]
 ~~($\Omega_{m0}$, $\beta$)~~ & ~~(0.272, $-0.02$)~~
  & ~~(0.308, $-0.02$)~~ & ~~(0.238, $-0.02$)~~ & ~~(0.272, 0.29)~~
  & ~~(0.272, $-0.22$)~~\\[0.7mm] \hline\\[-3.5mm]
 $|\dot{G}_{\rm eff}/G_{\rm eff}|$ & 0.919465 & 0.988932
  & 0.842703 & 9.74454 & 13.8775 \\[0.7mm] \hline\hline
 \end{tabular}
 \end{center}
 \caption{\label{tab2} The present value of
 $|\dot{G}_{\rm eff}/G_{\rm eff}|$ (in units of
 $10^{-12}\,{\rm yr}^{-1}$) with the best-fit parameters
 of~\cite{r22} and the corresponding $2\sigma$ edge for the
 case of $f(T)=-\mu T\left(1-e^{\beta T_0/T}\right)$.}
 \end{table}

%=================================================

%============================= section 3.3 ===================================

\subsection{$f(T)=-\mu T\left(1-e^{\beta T_0/T}\right)$}\label{sec3c}

\begin{center}
 \begin{figure}[tbp]
 \centering
 \includegraphics[width=0.5\textwidth]{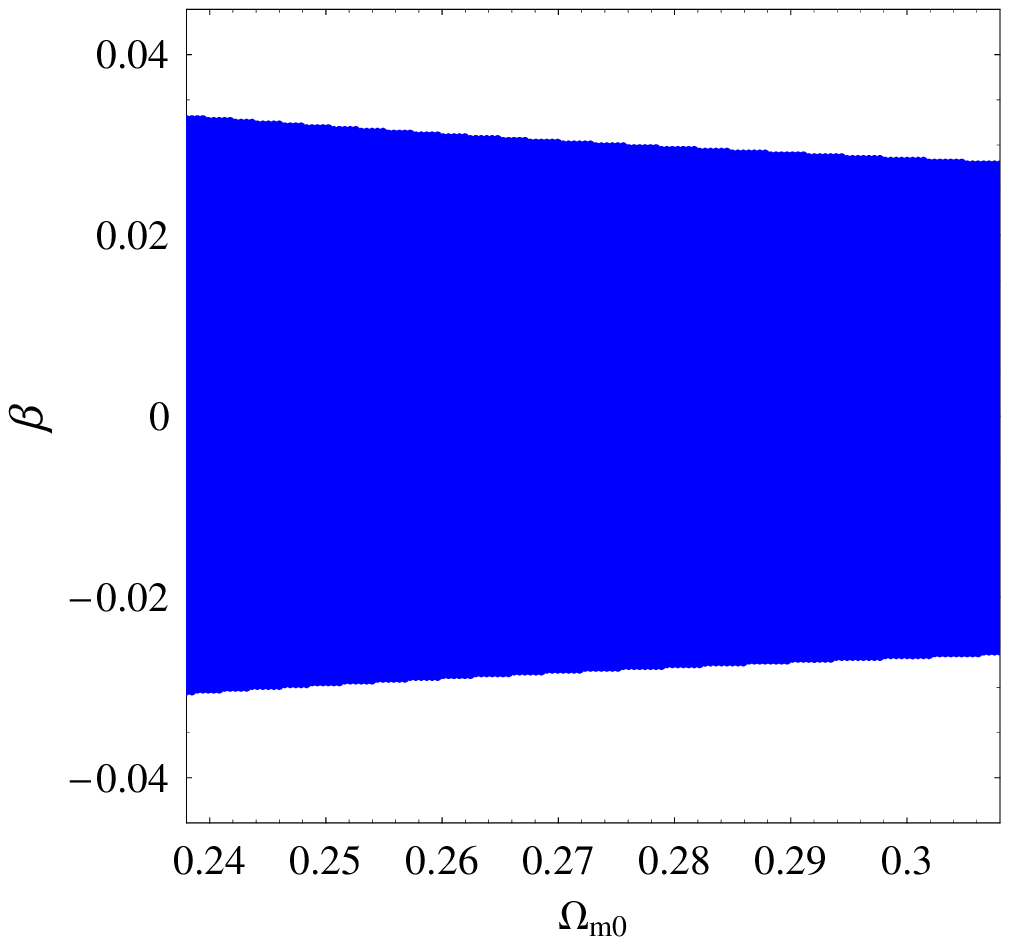}
 \caption{\label{fig2} The viable region in the
 $\Omega_{m0}-\beta$ parameter space which can satisfy the
 observational constraint on the varying gravitational
 ``constant'' given in Eq.~(\ref{eq1}), as well as the latest
 cosmological data (SNIa+BAO+CMB), for the case of
 $f(T)=-\mu T\left(1-e^{\beta T_0/T}\right)$. See text for
 details.}
 \end{figure}
 \end{center}

In this subsection, we consider the case of
 $f(T)=-\mu T\left(1-e^{\beta T_0/T}\right)$, where $\mu$
 and $\beta$ are both constants. Obviously, when $\beta\to 0$
 we have $f(T)\to \mu\beta T_0=const.$, so it reduces to
 $\Lambda$CDM model. Substituting
 $f(T)=-\mu T\left(1-e^{\beta T_0/T}\right)$ into the modified
 Friedmann equation~(\ref{eq19}), one easily finds that
 $\mu$ is not an independent model
 parameter~\cite{r21,r22,r31}, i.e.,
 \be{eq34}
 \mu=\frac{1-\Omega_{m0}}{1-\left(1-2\beta\right)e^\beta}\,.
 \ee
 It is easy to obtain
 \be{eq35}
 f_T=-\mu+\mu\left(1-\beta/E^2\right)e^{\beta/E^2}\,,~~~~~~~
 f_{T0}=-\mu+\mu\left(1-\beta\right)e^\beta\,,
 \ee
 \be{eq36}
 f_{TT}=\frac{\mu\beta^2}{T_0 E^6}\,e^{\beta/E^2}\,,~~~~~~~
 f_{TT0}=\frac{\mu\beta^2}{T_0}\,e^{\beta}\,,
 \ee
 where $E^2\equiv T/T_0=H^2/H_0^2$. On the other hand, substituting
 $f(T)=-\mu T\left(1-e^{\beta T_0/T}\right)$ into the modified
 Friedmann equation~(\ref{eq19}), we find that~\cite{r22,r31}
 \be{eq37}
 E^2=\Omega_{m0}(1+z)^3+\mu E^2\left[
 1-e^{\beta/E^2}+2\left(\frac{\beta}{E^2}\right)e^{\beta/E^2}
 \right]\,.
 \ee
 If $\beta\to 0$, we have $\mu\beta\to 1-\Omega_{m0}$ from
 Eq.~(\ref{eq34}), and hence Eq.~(\ref{eq37}) reduces to the
 one of $\Lambda$CDM model. Differentiating Eq.~(\ref{eq37})
 with respect to redshift $z$, we obtain
 \be{eq38}
 \frac{dE^2}{dz}=\frac{3\Omega_{m0}(1+z)^2}{1+\mu\left[\left(
 1-\beta/E^2+2\beta^2/E^4\right)e^{\beta/E^2}-1\right]}\,.
 \ee
 Substituting Eqs.~(\ref{eq35}), (\ref{eq36}) and~(\ref{eq38})
 into Eq.~(\ref{eq23}) and then Eq.~(\ref{eq22}), it is easy
 to see that at redshift $z=0$, we have
 \be{eq39}
 \left|\,\frac{\dot{G}_{\rm eff}}{G_{\rm eff}}\,\right|=
 \left|\,\frac{3\mu\beta^2 \Omega_{m0} e^\beta H_0}{\,\left[
 1-\mu+\mu(1-\beta)e^\beta\right]\left[1-\mu+\mu(1-\beta
 +2\beta^2)e^\beta\right]\,}\,\right|\,,
 \ee
 where $\mu$ have been given in Eq.~(\ref{eq34}). Therefore,
 if the model parameters $\Omega_{m0}$ and $\beta$ are given,
 we can correspondingly get the present value of
 $|\dot{G}_{\rm eff}/G_{\rm eff}|$ from Eq.~(\ref{eq39}).
 Note that in~\cite{r22}, this
 $f(T)=-\mu T\left(1-e^{\beta T_0/T}\right)$ theory has been
 constrained by using the latest cosmological data, i.e., 557
 Union2 type Ia supernovae (SNIa) dataset, baryon acoustic
 oscillation (BAO), and cosmic microwave background (CMB) data
 from WMAP7. The corresponding $2\sigma$ results are given
 by~\cite{r22}
 \be{eq40}
 \Omega_{m0}=0.272^{+0.036}_{-0.034}\,,~~~~~~~
 \beta=-0.02^{+0.31}_{-0.20}\,.
 \ee
 Again, we firstly try to see whether the present value
 of $|\dot{G}_{\rm eff}/G_{\rm eff}|$ with the best-fit
 parameters of~\cite{r22} and the corresponding $2\sigma$ edge
 can satisfy the observational constraint on the varying
 gravitational ``constant'' given in Eq.~(\ref{eq1}), namely,
 $|\dot{G}/G|\leq 1.3\times 10^{-12}\,{\rm yr}^{-1}$. From
 Table~\ref{tab2}, it is easy to see that the first three
 points (whose $\beta$ is close to zero) can satisfy the
 observational constraint on the varying gravitational ``constant''
 given in Eq.~(\ref{eq1}), whereas the last two points (whose
 $\beta$ is far away from zero) cannot. Also from Table~\ref{tab2},
 we find that for a fixed $\beta$, the smaller $\Omega_{m0}$,
 the smaller $|\dot{G}_{\rm eff}/G_{\rm eff}|$ is, whereas
 $|\dot{G}_{\rm eff}/G_{\rm eff}|$ is also somewhat correlated
 with the extent of $\beta$ deviating from zero. Next, we try
 to find the viable region in the $\Omega_{m0}-\beta$ parameter
 space which can satisfy the observational constraint on the
 varying gravitational ``constant'' given in Eq.~(\ref{eq1}),
 as well as the latest cosmological data (SNIa+BAO+CMB). To
 this end, we scan the $\Omega_{m0}-\beta$ parameter space
 within the $2\sigma$ region of~\cite{r22}, namely
 $0.238\leq\Omega_{m0}\leq 0.308$ and $-0.22\leq \beta\leq 0.29$
 (see Eq.(\ref{eq40})), and correspondingly calculate the
 present value of $|\dot{G}_{\rm eff}/G_{\rm eff}|$ at every
 scanned ($\Omega_{m0}$, $\beta$) point. The viable parameter
 region is determined by $|\dot{G}_{\rm eff}/G_{\rm eff}|\leq
 1.3\times 10^{-12}\,{\rm yr}^{-1}$, and we present it in
 Fig.~\ref{fig2}, which is approximately a trapezoid region. It
 is easy to see that the allowed $\Omega_{m0}$ is still
 unchanged, i.e., $0.238\leq\Omega_{m0}\leq 0.308$. However,
 the allowed $\beta$ is significantly shrunk from the
 original $-0.22\leq\beta\leq 0.29$ of~\cite{r22} to a very
 narrow range around zero, namely
 \be{eq41}
 -0.030 \;\lsim\; \beta \;\lsim\; 0.033 \,.
 \ee
 This is the additional constraint
 on $f(T)=-\mu T\left(1-e^{\beta T_0/T}\right)$ theory from the
 varying gravitational ``constant''. Clearly, this result also
 improves the previous constraint by an order of magnitude.

%============================= Fig. 2 =================================

%======================================================================

\vspace{-9mm}  % used here just for a comfortable typesetting

%============================= section 4 ===================================

\section{Conclusion}\label{sec4}

As is well known, a varying effective gravitational ``constant'' is
 one of the common features of most modified
 gravity theories~\cite{r2}. Of course, as a modified gravity
 theory, $f(T)$ theory is not an exception. Noting that the
 observational constraint on the varying gravitational ``constant''
 is very tight, in the present work we tried to constrain
 $f(T)$ theories with the varying gravitational ``constant''.
 We found that the allowed model parameter $n$ or $\beta$ has
 been significantly shrunk to a very narrow range around zero.
 In fact, the results improve the previous constraints by an
 order of magnitude.

%============================= acknowledgements ===================================

\section*{ACKNOWLEDGEMENTS}
We thank the anonymous referee for quite useful comments and
 suggestions, which help us to improve this work. We are
 grateful to Professors Rong-Gen~Cai and Shuang~Nan~Zhang
 for helpful discussions. We also thank Minzi~Feng, as well
 as Puxun~Wu, Yi-Fu~Cai, Rong-Jia~Yang, Qing-Guo~Huang,
 Xiao-Jiao~Guo and Long-Fei~Wang, for kind help
 and discussions. This work was supported in part by NSFC
 under Grants No.~11175016 and No.~10905005, as well as NCET
 under Grant No.~NCET-11-0790, and the Fundamental Research
 Fund of Beijing Institute of Technology.

\renewcommand{\baselinestretch}{1.1}

%============================= references ==================================

\end{document}